\documentclass[12pt]{article}
\usepackage{graphicx}
\usepackage{pdfpages}
\usepackage{hyperref}
\usepackage{amssymb}
\usepackage{color}
\usepackage[normalem]{ulem}

\def\text#1{\mbox{#1}}

\newcommand{\be}{\begin{equation}}
\newcommand{\ee}{\end{equation}}
\newcommand{\beq}{\begin{eqnarray}}
\newcommand{\eeq}{\end{eqnarray}}

\makeatletter
\@addtoreset{equation}{section}

\makeatletter
\renewcommand\section{\@startsection {section}{1}{\z@}%
                                   {-3.5ex \@plus -1ex \@minus -.2ex}%nn
                                   {2.3ex \@plus.2ex}%
                                   {\normalfont\large\bfseries}}
\renewcommand\subsection{\@startsection{subsection}{2}{\z@}%
                                     {-3.25ex\@plus -1ex \@minus -.2ex}%
                                     {1.5ex \@plus .2ex}%
                                     {\normalfont\bfseries}}

\def\Re{{\rm Re}}
\def\Im{{\rm Im}}

\begin{document}

\begin{center}
{\Large \bf{A Worldsheet Description of \\\vspace{1mm} Instant Folded Strings }}

\vspace{10mm}

\textit{Akikazu Hashimoto${}^{(1)}$,   Nissan Itzhaki${}^{(2)}{}^{(3)}$ and Uri Peleg${}^{(2)}$}
\break \break
${}^{(1)}$ Department of Physics, University of Wisconsin, Madison, WI 53706, USA \\
${}^{(2)}$ School of Physics and Astronomy, Tel Aviv University, Ramat Aviv, 69978, Israel \\
${}^{(3)}$ Institute for Advanced Study, Princeton, NJ 08540\\

\begin{abstract}

\noindent Time-like linear dilaton backgrounds admit a classical solution that describes
a closed folded string that is created at an instant. We refer to such strings as Instant Folded Strings (IFS). We study an exact worldsheet CFT description of an IFS that involves two vertex operators which describe two open string modes that propagate on a time-like FZZT-brane, which plays the role of a regulator to the IFS. We take advantage of this description to calculate the most basic quantity associated with IFSs - their production rate. Some implications of this calculation to stringy cosmology and black hole interior are briefly discussed.

\end{abstract}

\end{center}

\newpage

\baselineskip18pt

\section{Introduction}

In order to understand  cosmology and the black hole interior in  string theory we have to understand how to formulate string theory in time dependent situations. This was proven to be a challenging task
that is likely to involve some non-trivial conceptual development (for a recent discussion see \cite{WittenTalk}).

\begin{figure}
\centerline{\includegraphics{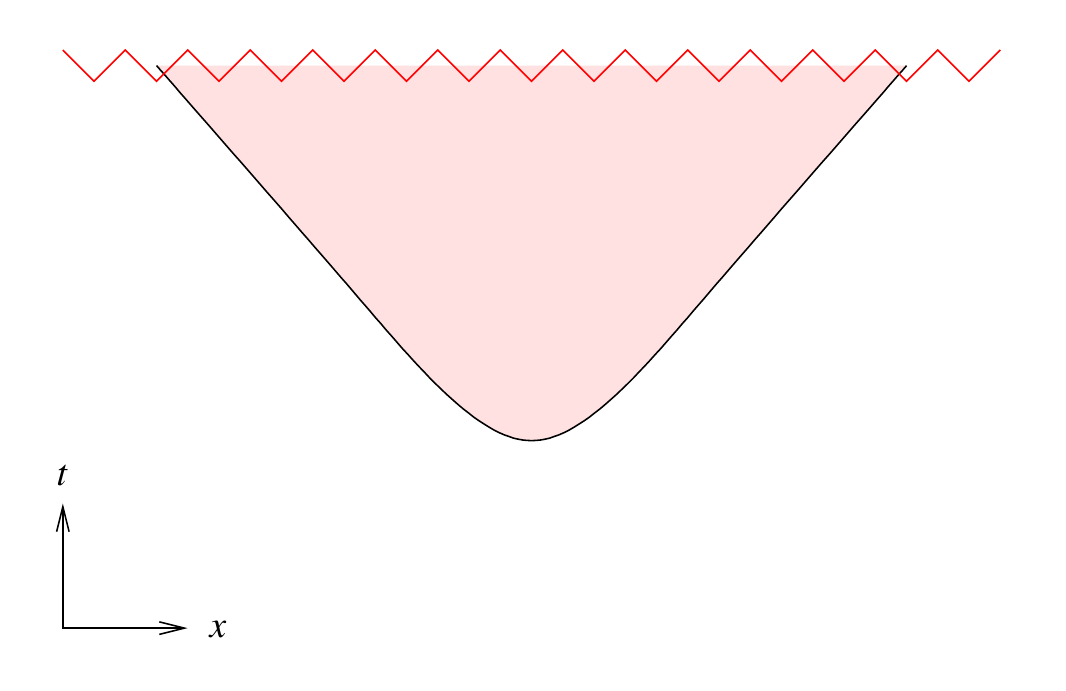}}
\caption{The red zigzag line marks the space-like singularity due to the time-like linear dilaton.  This background admits an unconventional string solution - a closed folded string that is created {\it classically} at a certain point in space-time. \label{figifs}}
\end{figure}

 Already  the simplest   time dependent background, time-like linear dilaton (for which there is an exact CFT description),  appears to involve non standard physics.  Ordinary closed string modes, that propagate from past infinity to future infinity are described on the worldsheet,  as usual, via local $(1,1)$ operators. However, on top of these, there are classical solutions  that describe closed folded strings  that, for positive dilaton slope, appear out of the vacuum at an instant and propagate to  future infinity while expanding at the speed of light \cite{Itzhaki:2018glf} (see figure (\ref{figifs})).   We refer to such strings as Instant Folded Strings (IFS).

IFSs are not described by local $(1,1)$ operators and
so far  were studied only at the classical level. This study revealed some potentially fascinating properties. For example a single IFS violates the averaged null energy condition \cite{Attali:2018goq} and as a result IFSs could induce negative pressure at no energy cost in cosmological setups \cite{Itzhaki:2021scf}. This suggests that  IFSs might play an important role in black hole physics and in Cosmology.  To explore this we have to understand IFSs beyond the classical limit. What we need is to  learn how to incorporate IFSs into the textbook string theory formalism.  In particular, we should be able  to calculate, from first principles, the creation rate of IFSs as well as their interactions with standard stringy modes and among themselves.

To this end we have to find their exact CFT description.
The approach presented here is related to \cite{Maldacena:2005hi} via Wick rotation. In \cite{Maldacena:2005hi} it was pointed out that in space-like linear dilaton backgrounds there are classical solutions that describe a Long Folded String (LFS) that are stretched all the way to the weak coupling. Moreover, it was realized that a LFS can be regularized with the help of a FZZT-brane. This provides an exact CFT description of a (regularized) LFS in terms of {\em two} vertex operators that describe an ingoing and an outgoing open string modes that propagate on the FZZT-brane.  The LFS appears as an intermediate state in this scattering process. In the next section we review this work.

In section 3 we Wick rotate the setup of \cite{Maldacena:2005hi} and find the exact CFT description of an IFS in terms of two vertex operators that describe two outgoing open string modes that propagate on a time-like FZZT-brane.
In section 4 we use this exact CFT description of an IFS to calculate its simplest property - its creation rate. In section 5 we briefly discuss some implications of this calculation to the interior of near extremal black NS5-branes and to cosmology that is driven by time-dependent dilaton.

In Appendix \ref{appA}, we provide an extended account of the calculation of FZZT open string two point amplitude for the space-like and the time-like Liouville theory. The reason we present this computation at this level of detail is that the dynamics of FZZT amplitude in time-like Liouville theory turns out to be extremely subtle. 
 There are several previous works on FZZT  branes in time-like Liouville theory \cite{Gutperle:2003xf,Bautista:2021ogd},  these works did not seem to fully address the issues we encountered. 
Fortunately, we found that all of these subtleties can be attributed to the FZZT branes serving as a regulator to the IFS, and that the IFS part of the dynamics can be extracted cleanly, independent of the subtleties of the FZZT brane dynamics.  
We did not resolve all of the issues, and one possible conclusion is that FZZT brane in time-like Liouville theory  is  not a consistent physical system and can only be viewed as a regulator. We believe it is somewhat premature to settle on that conclusion, and instead hope that our analysis can provide the foundation for further study of FZZT branes in time-like Liouville theory. Since Appendix \ref{appA} is somewhat long, we provided the summary of our technical findings in section \ref{appAsummary}.

\section{LFSs and FZZT-branes}

In this section we review the construction of LFS and its regularization in terms of FZZT-brane \cite{Fateev:2000ik,Teschner:2000md}. While \cite{Maldacena:2005hi} focused on $b=1$, that is relevant for the $c=1$ matrix model, we consider a general space-like linear background
as we wish to Wick rotate $b$ in the next section.

The relevant part of the background is
\be
ds^2=-dt^2+d\phi^2,~~~\Phi=Q\phi,
\ee
which describes a space-like linear dilaton, $\phi$ and a time direction,
$t$. We take $Q>0$ which means that strong coupling is at large and positive values of $\phi$. The full background in string theory can involve other compact and non-compact directions that render the theory critical.\footnote{Strictly speaking, for $Q > 2$, the central charge of ${\cal M}$ needs to be negative in order for the target space-time to be critical, but we will not worry too much about this issue. In the time-like case, which is the subject of primary interest in this article, sectors with negative central charge do not appear.}

Strings on this background admit LFS solutions \cite{Maldacena:2005hi} that take the form
\be\label{lfs}
t=\sigma^0, ~~~\phi=\phi_0-Q\log\left( 1/2 \left(\cosh(\sigma^0/Q)+\cosh(\sigma^1/Q)\right)    \right) \ .
\ee
Unlike in standard string solutions, the range of the coordinates is $-\infty < \sigma^0,~ \sigma^1 <\infty$. This implies that the exact CFT description associated with (\ref{lfs}) is not in terms of a $(1,1)$ operator.
The solution describes an infinite  LFS that is stretched all the way to the weak coupling region ($\phi=\infty$). The point where the string folds ($\sigma^1=0$) moves from $\phi=-\infty$ to $\phi=\phi_0$ and  back to $\phi=-\infty$. $\phi_0$ can be chosen such that the entire LFS is in the weakly coupled region.

The energy associated with (\ref{lfs}) is infinite and as was discussed in \cite{Maldacena:2005hi} it can be regularized with the help of FZZT-branes.
The worldsheet action that describes the linear dilaton is
\be S =  \int d^2 \sigma \left({1 \over 4 \pi} (\partial \phi)^2 + {1 \over 4 \pi} Q R \phi \right) , \qquad Q = b + {1 \over b} \label{spaceSq} \ee
with central charge
\be c = 1 + 6 Q^2 \label{centralcharge}.\ee
Adding an FZZT-brane amounts to adding to (\ref{spaceSq}) the truly marginal boundary deformation \cite{Fateev:2000ik,Teschner:2000md}
\be\label{bbo}
\int d \sigma \, \lambda e^{b \phi}.
\ee
The FZZT-brane is located in the region in which the boundary term  (\ref{bbo}) is small. Namely, it is stretched from $\phi=-\infty$ to
\be\label{loc}
\phi_{FZZT}\sim \log(\lambda)/b.
\ee
Much like $\phi_0$, $\lambda$ can be chosen such that the FZZT-brane is located entirely at weak coupling.

The boundary vertex operator is
\be B_\beta(\sigma)= e^{\beta \phi(\sigma)},\ee
with dimension
 \be
\Delta_\beta = \beta(Q-\beta)  \label{deltabeta}.\ee
Physical states, associated with open strings that live on the FZZT-brane, are parameterized by
\be \beta = {Q \over 2} + i P \label{beta},\ee
where $P$ is the momentum of the open string.

\begin{figure}
\centerline{\includegraphics{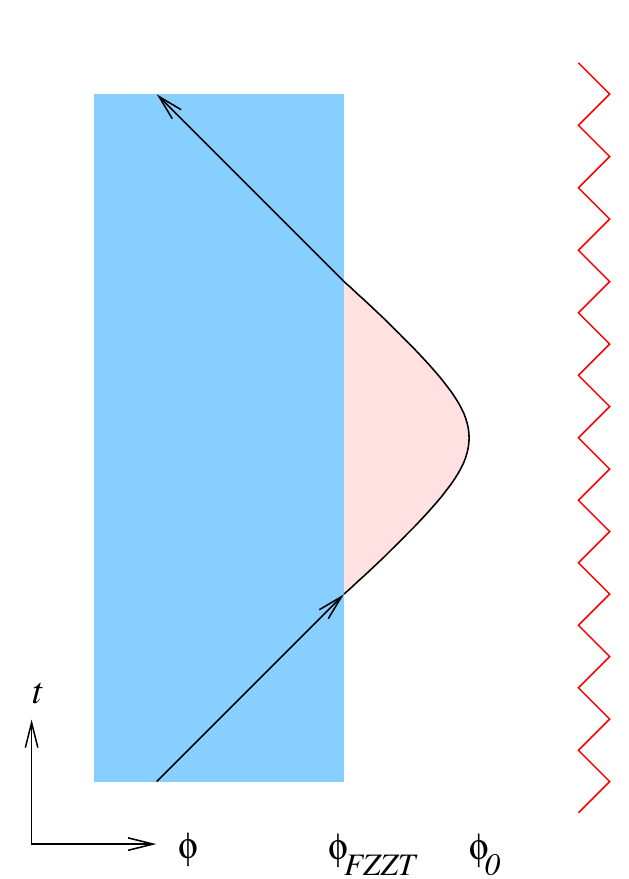}}
\caption{Long folded strings on FZZT branes. The region $\phi < \phi_{FZZT}$ where the boundary Liouville potential is small is illustrated in blue. The long string stretches from $\phi_{FZZT}$ to $\phi_0$. \label{figc}}
\end{figure}

Consider  an  open string with  $P>0$ and $E=P$ that lives on  the FZZT-brane\footnote{We focus on energetic modes for which the on-shell condition implies that, to a good approximation,  $E=|P|$.} and propagates from the weakly coupled region towards strong coupling. We take $\phi_{FZZT}$  such that long before strong coupling the brane ends. When the open string mode reaches the end of the brane, $\phi_{FZZT}$, its momentum causes the string  to fold and be stretched away from $\phi_{FZZT}$ while keeping its end points ($\sigma^1=0,~\pi$) at $\phi_{FZZT}$ (see figure \ref{figc}). Momentum conservation implies that the string will reach
\be
\phi_0=\phi_{FZZT} + \frac{P}{2T},
\ee
where $T$ is the string tension and the factor of $2$ is due to the fact that the string is folded. Its tension causes the LFS to shrink towards the FZZT-brane until eventually it becomes a mode with $E=-P>0$  that propagates back toward weak coupling.

When $E\gg M_s$ the size of the folded string that is stretched from the FZZT-brane is large and its shape is that of the LFS (\ref{lfs}) with a cutoff at $\phi_{FZZT}$ (see figure \ref{figc}).  So from the point of view of the FZZT-brane a regularized LFS is an intermediate state in a  pure reflection  scattering process that  is described by a two point function of open strings. One with $E=P>0$ and one with $E=-P>0$.
In the standard  convention, in which all modes are ingoing, the relevant two point function on the FZZT-brane is
\be d(\beta) = \langle B_{\beta}(\sigma) B_{\beta}(\sigma') \rangle (\sigma-\sigma')^{2 \Delta_\beta} \ , ~~~~\mbox{with}~~~~P>0.\ee

The two open string vertex operators provide the exact CFT description of a regularized LFT that reaches $\phi_0$. Taking the cutoff to infinity amounts to taking
\be
P\to\infty, ~~~~ \phi_{FZZT}\to - \infty
\ee
while keeping $\phi_0$ intact.

To establish this picture we take advantage of the exact two point function on FZZT-brane,
\be d(\beta) = \langle B_{\beta}(\sigma) B_{\beta}(\sigma') \rangle (\sigma-\sigma')^{2 \Delta_\beta} \ ,  \ee
that was calculated in \cite{Fateev:2000ik,Teschner:2000md}. The details of this calculation are presented in the appendix. Here we focus on  the large $P$ limit in which the phase shift associated
with this pure reflection  process takes the form
\beq
\log d(\beta) & = &  \langle e^{(Q/2 + i P) \phi(\sigma)} e^{(Q/2 + i P) \phi(\sigma')} \rangle (\sigma -\sigma')^{2 \Delta_\beta} \cr
&=&
 i \pi  P^2+2 i Q P \log (Q P)  \label{amp} \\
&& + \left(2 i  Q (\log ( 2Q^{-1}  )-1)-{2 i  \over b}\log \left({2\pi \lambda b^{(1-b^2)} \over \Gamma(1-b^2)}\right) \right) P
+{\cal O}  (P^0)
. \nonumber \eeq
There are three contributions to this phase:

\begin{itemize}
\item[(1)] The term
\be  i\pi P^2\ee
is associated with the bulk of the LFS. Namely, the Nambu-Goto action gives (in units where $\alpha^{'}=1$)
\be
S_{NG}=\frac{A}{\pi}=\frac{L^2}{\pi}=\pi P^2,
\ee
where we used  the fact that  $L=\pi P$.\\
\item[(2)] The term
\be\label{ftf} 2i QP\log(QP) \ee
is associated with the fold of the LFS. To see this we note that the fold, $\sigma^1=0$, is located at
\be\label{one}
\phi(\sigma^0)=-Q\log(1+\cosh(\sigma^0/Q))
\ee
which satisfies
\be
\ddot{\phi}(\sigma^0)+e^{\phi(\sigma^0)/Q}=0.
\ee
Namely, it corresponds to a particle moving in a potential
\be
V(\phi)=Q e^{\phi/Q}.
\ee
Indeed  (\ref{ftf}) is the phase shift associated with high momentum scattering of this potential.

\item[(3)] Finally, there is a  term linear in $P$
\be \left(2 i  Q (\log ( 2Q^{-1}  )-1)-{2 i  \over b}\log \left({2\pi \lambda b^{(1-b^2)} \over \Gamma(1-b^2)}\right) \right) P \sim - 2 i  \phi_{FZZT} P \ , \ee
which defines the quantity $\phi_{FZZT}$.
This term is associated with the propagation of the mode on the FZZT-brane.  Note that in the small $b$ limit,
$ \lambda e^ {b\phi_{FZZT}}  \sim  1 $
in agreement with (\ref{loc}).
\end{itemize}

Therefore, (1) and (2) above are attributed to the LFS physics while (3) is to the FZZT-brane regulator. The fact that one can separate, at large $P$,  between the system one wishes to describe and the regulator implies that this is a valid regulator. We conclude, therefore, that
\be \log \tilde A' =  i\pi P^2 +  2i QP\log(QP) \ ,\label{logA} \ee
is the phase associated with the LFS.

\section{IFS and time-like FZZT-branes}

In this section we consider the Wick rotation of the discussion in the previous section.
The action and  vertex operators associated with the time-like case is obtained from the space-like case via an analytic continuation
\be b = - i \bar b, \qquad \phi = i \bar \phi, \qquad Q = i q \label{continue}. \ee
For example, the relevant part of the background is
\be
ds^2=-d\bar\phi^2+dx^2,~~~\Phi=-q \bar\phi,
\ee
 where  now $\bar \phi$ is a time-like direction and, as in the previous section, the full background  involves other compact and non-compact directions that render the theory critical.  We focus on the case where $q<0$ in which the string coupling blows up in the future.

The worldsheet action that describes $\bar\phi$ is
\be S =  \int d^2 \sigma \left(-{1 \over 4 \pi} (\partial \bar  \phi)^2 - {1 \over 4 \pi} q R \bar \phi \right) + \int d \sigma\,  \lambda e^{\bar b \bar \phi}, \qquad q = - \bar b + {1 \over \bar b} \ , \label{TLTaction}\ee
with central charge
\be c = 1  - 6 q^2. \ee
The region where the boundary Liouville potential is small is $\bar \phi \ll 0$.

\begin{figure}
\centerline{\includegraphics{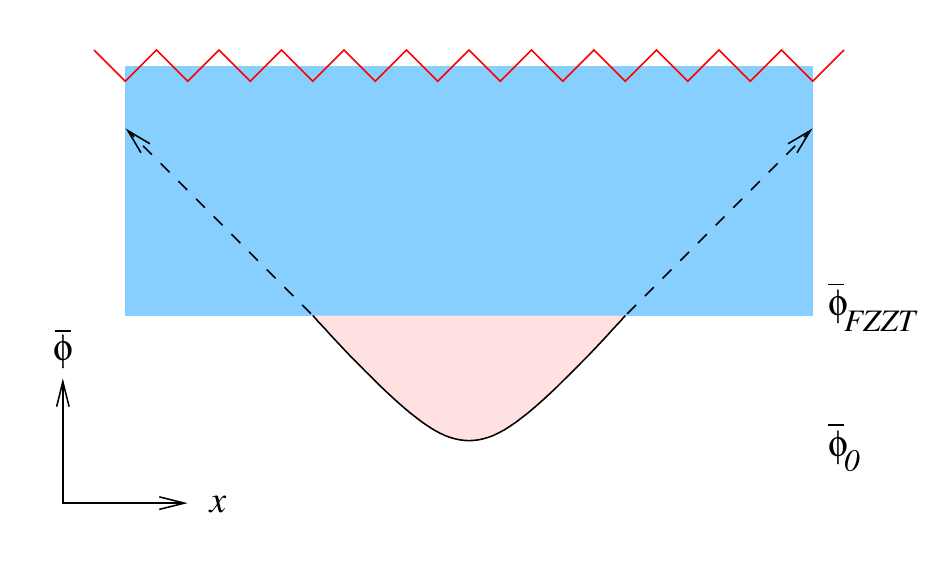}}
\caption{Schematic illustration of timelike FZZT branes and the folded strings with $-1 <\bar b < 0$.  The blue shaded region describes the FZZT-brane. 
For these values of $\bar b$, the negative energy open string states (indicated by dotted arrow)  propagate in the region where the Liouville potential is small (and the string coupling is strong.)  \label{figd}}
\end{figure}

The Wick rotation of the LFS is an IFS that takes the form
\be\label{ifss}
x=x_0+\sigma^1,~~~~\bar\phi=\bar\phi_0 - q\log\left( 1/2 \left(\cosh(\sigma^0/q)+\cosh(\sigma^1/q)\right)    \right) \ .
\ee
This solution
describes a closed folded string that is created at $x=x_0$ and at time $\bar\phi_0$. The points where the string folds, $\sigma^1 = 0$, are moving faster than light. At the creation point the speed is infinite and it approaches the speed of light at later times. See figure \ref{figd}.

A crucial difference between the LFS and the IFS is that while a LFS costs an infinite amount of energy, an IFS is massless. Hence IFSs can dramatically affect the IR physics.  An IFS is massless since it is created classically from the vacuum.
It is massless in an interesting way that is quite relevant for its open string description. The bulk of the IFS has positive energy density due to the tension of the folded string. This positive energy is canceled against the negative energy at the fold. Since the size of the IFS grows with time the negative energy at the fold also grows with time. As discussed below the open string vertex operator that  provide the CFT description of the IFS are determined  by the energy at the IFS fold.

In order to regulate the IFS as an FZZT amplitude, we need to also analytically continue the vertex operator. We see that
under\footnote{We are using the convention that $P$ and $\bar P$  are real and positive in their respective contexts.}
\be P = i \bar P \ee
we get
\be B_\beta = e^{(Q/2 + i P) \phi} = e^{-(q/2  +i \bar P) \bar \phi} =e^{-\bar \beta \bar \phi}\equiv B_{\bar \beta} \ee
and so we define
\be \bar \beta \equiv  (q/2 + i \bar P) = - i \beta.\ee

With these conventions, we can immediately consider the analytic continuation of (\ref{amp}) as\footnote{Strictly speaking, this is a dangerous thing to do since  Wick rotation of an approximation might lead to an answer that is far from the actual result. What should  be done is to first analytically continue the {\em exact} expression of the two point function   and only then take the large $\bar P$ limit. This is done in the appendix. }
\beq \log D_{\bar b}(\bar \beta) & = & \log(\langle e^{(-q/2- i \bar P) \bar \phi}(\sigma)  e^{(-q/2 - i\bar P) \bar \phi}(\sigma')\rangle (\sigma - \sigma')^{2 \Delta_{\bar \beta}}) \cr
& = &
  i \pi \bar P^2- 2 i q\bar P \log(  q\bar P)  \label{logD3} \\
 &&- \left(2 i  q (\log ( 2 \bar b q^{-1} )-1)-{2 i  \over \bar b}\log \left({-2\pi \lambda \bar b^2 \over \Gamma(1+\bar b^2)}\right) \right)\bar P
+{\cal O}  (\bar  P^0) \nonumber \eeq
where we take $-1< \bar b < 0$ so that $q < 0$.  This is the process illstrated in figure \ref{figd}.

Note that in figure \ref{figd}, the FZZT brane occupies the future region so that the open strings are outgoing, but the sign of the $i \bar P \bar \phi$ in the exponential of the vertex operator in (\ref{logD3}) is negative. We take this to mean that the open strings are carrying negative energy. This is consistent with the fact that the energy carried by the fold of the IFS is negative. Therefore, we must interpret the FZZT brane in \ref{figd} as a negative brane in the sense of \cite{Okuda:2006fb,Dijkgraaf:2016lym}.

We see, just as in the spatial case (\ref{logA}), that the terms characterizing the IFS in (\ref{logD3})
\be \log \tilde A' =  + i \pi  \bar P^2-2 i q \bar P  \log(q \bar P)  \label{logA2} \ee
are purely imaginary.  Note the minus sign in the second term which reflects the fact that the energy at the fold of an IFS is negative, as opposed to the LFS where it is positive. This fact will be used in the following section. 

The status of the term of order $\bar P$ is somewhat subtle. Specifically, this term is not manifestly purely imaginary. We interpret this subtly as reflecting a pathology of the physics of open string dynamics on the FZZT brane which will not affect the discussion of the IFS production rate below. There are also subtleties in this expression at order $\bar P^0$. We will elaborate further on these subtleties in the appendix.

In summary, we are taking the limit
\be\label{bg}
\bar  P\to\infty,~~~\bar\phi_{FZZT}\to\infty,
\ee
while keeping
\be\label{kl}
\bar\phi_0= \bar\phi_{FZZT}-\frac{\bar P}{2 T},
\ee
fixed, as illustrated in figure \ref{figd}, in order to extract from it the amplitude (\ref{logA2}), which is associated with the IFS production process.

\section{IFS production rate \label{sec:rate}}

In this section we use the exact CFT description of the IFS in terms of the two  vertex operators $V_1$ and $V_2$ associated with the open strings that propagate on the time-like FZZT-brane (\ref{logD3}) to calculate the IFS production rate.

We consider a space-time of the form $R^{\bar \phi} \times R^x \times {\cal M}$ where ${\cal M}$ is a  manifold whose central charge is $c_{\cal M} = 24 + q^2$ and $R^{\bar \phi}$ is the time-like linear dilaton direction. The vertex operator we  consider does not involve ${\cal M}$ and  takes the form
\be  V_i(\sigma_i) = e^{-q/2-i \bar P_i \phi(\sigma_i) + i p_i x(\sigma_i)}\ , \ee
where we take $\bar P$ to parameterize the energy in the $\bar \phi$ direction and $p$ to label the momentum along $x$. We will be considering an open string tachyon whose on-shell condition is
\be -{q^2 \over 4} - \bar P_i^2  + p_i^2 = 1, \qquad i =1,2 \ . \ee
The open string amplitude that is relevant for the IFS production rate is the two-point function
\be
A(\bar P)=\langle V_1(\sigma_1) V_2(\sigma_2)\rangle  \ ,
\ee
with kinematics
$ \bar P \equiv \bar P_1 = \bar P_2 $ and $ p \equiv p_1 = - p_2  \ . $

As usual  we can use two of the three ghost zero modes to locate $V_1$ at, say, $z_1=0$ and $V_2$ at $z_2=\infty$.  This leaves
a residual conformal group factor, $\mbox{Vol(res)}$, associated with dilatation which keeps
$\sigma_1=0$ and  $\sigma_2=\infty$ intact. Hence
\be\label{sta}
A(\bar P)=\frac{1}{\mbox{Vol(res)}}\delta(\bar P_1-\bar P_2)\delta(p_1+p_2)\tilde{A}(\bar P),
\ee
where  $\tilde{A}(\bar P)$ is the amplitude (\ref{logD3}).

 Since $p_1=p=-p_2$ the $\delta(p_1+p_2)$ yields an infinity that is related to the infinite size of the  spacial direction. Concretely,
\be
\delta(p_1+p_2) \to \frac{L}{2\pi},
\ee
which, as usual, reflects the fact that the amplitude describes a process -  in our case, an IFS creation  - that can take place anywhere in the $x$ direction.

The $\delta(\bar P_1-\bar P_2)$ is more interesting. Since $\bar P_1=\bar P_2=\bar P$ it diverges, which reflects the fact that the amplitude is invariant under an overall shifting of the $\bar\phi$ coordinate. Namely, shifting the time at which the IFS is created, $\bar \phi_0$, as well as the time at which the FZZT-brane is created, $\bar \phi_{FZZT}$,  while keeping $\bar P$ (and the size of the IFS) intact. This divergence is cancelled against the infinity of
$\mbox{Vol(res)}$.  To see this we parameterize the  residual conformal group in the standard fashion
\be\label{goku}
\tilde \sigma =  \exp(\alpha) \sigma,
\ee
and recall that in a linear dilaton theory (\ref{goku})
$\bar\phi(\sigma, \bar \sigma)$ varies under coordinate change, $\sigma\to \tilde \sigma(\sigma)$, in the following way  \cite{Polyakov:1981rd}
\be
\phi(\tilde \sigma, \bar {\tilde \sigma})=\phi(\sigma, \bar \sigma)-\frac{q}{2} \log|d \tilde \sigma/d\sigma|^2.
\ee
Therefore, (\ref{goku})
 induces a shift
\be\label{gohan}
 \bar\phi \to \bar\phi + q\Re({\alpha}).
\ee
Namely, just like $\delta(\bar P_1-\bar P_2)$, this shifts the whole process in the time direction (both FZZT-brane and IFS).

This means that, just like in \cite{Erbin:2019uiz}, a shift in $\bar\phi$ can be absorbed by a shift in $\alpha$
\be
\bar\phi\to \bar\phi+\epsilon,~~~\Re({\alpha})\to \Re({\alpha})+\epsilon q,
\ee
which implies that
\be
\frac{\delta(\bar P_1-\bar P_2)}{\mbox{Vol(res)}} = \frac{\frac{1}{2\pi}\int \bar\phi}{\int d^2\alpha} = \frac{\frac{1}{2\pi}\int \bar\phi}{\int d\Im(\alpha)d\Re(\alpha)} = \frac{q}{(2\pi)^2} \label{div},
\ee
where the additional factor of $2\pi$ in $\mbox{Vol(res)}$  is due to the integration range over the imaginary component. Thus, (\ref{sta}) gives
\be
A(p_1,p_2)= ~ \frac{q}{(2\pi)^2} ~ \tilde{A}_p ~ \delta(p_1+p_2). \label{amp2p}
\ee

This amplitude  describes the  creation from the vacuum of two open string modes that propagate on the FZZT-brane.  More precisely, as a function of time, $\bar \phi$, the process described by (\ref{amp2p}) is
\be
|0\rangle \to | IFS \rangle \to |p_1, p_2 \rangle.
\ee
The first transition takes place at $\bar\phi_0$ and the second at $\bar\phi_{FZZT}$.
To find  the production rate of IFS we need
to separate the first transition from the second. This is done by cutting the amplitude at
$\bar \phi_{FZZT}$ which gives a factor of $1/g_{FZZT}=\exp(-\bar \phi_{FZZT})$ due to the topology change associated with this  cutting. Namely,
\be\label{lq}
A_{IFS}(\bar\phi_{FZZT})= \frac{q}{(2\pi)^2 g_{FZZT}} ~ \tilde A'_p ~ \delta(p_1+p_2),
\ee
where now $p_1$ and $p_2$ are associated with the momentum at the  left and right folds of the IFS at $\bar \phi_{FZZT}$ and $\tilde A'_p $ is  (\ref{logA2}) computed in the previous section.

The way we interpret the $1/g_{FZZT}$ factor in (\ref{lq})  is the following.  The amplitude associated with the creation of an IFS at $\bar\phi_0$ is
\be
A_{IFS}(\bar\phi_{0})= \frac{q}{(2\pi)^2 g_0} ~ \tilde{A}'_p ~ \delta (p_1+p_2)~~~~\mbox{with}~~~~g_0=\exp(\bar \phi_{0}),
\ee
and that during the time evolution from $\bar\phi_0$ to
$\bar \phi_{FZZT}$ the IFS's wave function picks up a factor of $g_0/g_{FZZT}$ due to the fact that  a worldsheet field, $F$, and a canonically normalized field in space-time, $\tilde{F}$, are related via
$
F=g_s \tilde{F}.
$

The probability  of creating, from the vacuum, an IFS with $p_1$ and $p_2$ is, therefore
\be
d {\cal P}=|A_{IFS}(\bar\phi_{0})|^2\ dp_1 \, d p_2 =\frac{q^2 L}{(2\pi)^5 g_0^2} ~ \delta(p_1+p_2) \ dp_1 \, d p_2.
\ee
 Here, we used the fact that $|\tilde A'|^2 = 1$ since $\tilde A'$ given in (\ref{logA2}) is pure phase.
The $\delta (p_1+p_2)$  implies that $p_1=p=-p_2$ and so the probability of creating an IFS  at an interval $dx$ with momentum between $p$ and $p+dp$ is
\be
d{\cal P} = \frac{q^2}{(2\pi)^5 g_0^2} ~ dx\, dp .
\ee
At large $\bar P$, the on-shell condition implies that $dp=d\bar P$, and since
 (\ref{kl}) implies that   $d \bar\phi_0=\pi d\bar P$ we conclude that
$d \bar\phi_0=\pi p$. This means that
\be
\Gamma_{IFS} = \frac{d{\cal P}}{dx\,  d \bar\phi}=C\frac{q^2}{g_0^2},~~~\mbox{with}~~~C=\frac{1}{32\pi^6}
\ee
is the IFS production rate.

\section{Discussion}

The aim of this paper was to provide a CFT description of an IFS that so far was considered  only at the classical level.
In the spirit of \cite{Maldacena:2005hi} the description involved two open string modes that live on a time-like FZZT-brane which plays the role of a regulator. Using this description we calculated the simplest quantity associated with IFSs - their production rate.

The validity of this calculation, which was done using the exact time-like linear dilaton CFT  is expected to be more general.
The reason is that the creation of an IFS is a local  process that takes place around its tip, ($x=x_0$ and $\bar\phi= \bar\phi_0$ in (\ref{ifss})).  The characteristic length associated with IFS creation 
scales like $\sim Q$, and for small $Q$, it is much smaller than the scale associated with the background which scales like, $ 1/Q$. Hence it is natural to expect that IFS are created whenever the string coupling grows with time, and that
\be\label{jh}
\Gamma_{IFS} = C  \dot \Phi_0^2 e^{-2\Phi_0},
\ee
is a good approximation to the IFS  production rate when $0<\dot\Phi_0=\frac{d\Phi}{dt}|_{\bar\phi=\bar\phi_0} \ll 1$ and the curvature is small.  Similarly, (\ref{ifss}) is expected to be a good approximation to the IFS solutions at time scales that are much shorter than the time scale associated with the curvature of the background (and $\ddot\Phi$).

We would like to briefly discuss now possible implications of (\ref{jh}) to the black hole interior in string theory and to stringy cosmology.

\begin{itemize}
\item {\it Black NS5-branes interior} \\

The region behind the horizon of a  black hole associated with  $k$ NS5-branes, with $k\gg 1$, is such that $0<\dot\Phi \ll 1$\footnote{As long as we are not too close to the singularity.} and so IFS are expected to be created approximately at the rate (\ref{jh}). This means that the black NS5-branes interior is not empty, rather it is filled with IFS. Calculating their backreaction in a generic state is a difficult task\footnote{For the specific state in which they are all created at the horizon bifurcation this was done in \cite{Giveon:2020xxh}.}. However, even without taking into account  the IFSs backreaction the following simple argument suggests that their production rate (\ref{jh}) scales with $k$ and the string coupling constant  in just the right way to support the claim made in \cite{Itzhaki:2018glf,Attali:2018goq,Giveon:2019gfk,Giveon:2019twx,Jafferis:2021ywg,Giveon:2020xxh} that IFSs provide the microscopic description of black NS5-branes.

To see this we consider an infalling observer, and ask how many IFSs such an observer encounters along his/her trajectory towards the singularity.  Since the folds of the IFS are moving in nearly lightlike trajectories, this question is reduced to asking how many IFSs are produced within
the causal diamond connecting the horizon bifurcation point and the  point where the observer hits the  singularity (the blue
region in figure 4).
Since the size of the blue region in figure 4 scales like $k$, and since  typically behind the horizon $\dot\Phi\sim 1/\sqrt{k}$ it is easy to see from (\ref{jh}) that the number of IFS encounters by an infalling observer indeed scales like the black hole entropy
$
 \frac{1}{g^2}\sim S_{BH}.
$

This could have implications to a Schwarzschild black hole as well. Reducing the sphere in the  Schwarzschild geometry one obtains approximately an $SL(2)/U(1)$ black hole.\footnote{See  e.g. \cite{Emparan:2013xia} for the validity of this approximation as a function of dimension.} In particular, $0<\dot\Phi$ behind the horizon is replaced by the fact that the size of the sphere shrinks with time behind the horizon of a Schwarzschild black hole. If this triggers the creation of  IFS (presumably in S-wave) then the analog of (\ref{jh}) and the discussion above might support the speculation made in
\cite{Giveon:2021gsc} that
IFSs provide the microscopic description also of Schwarzschild black hole.

\begin{figure}
\centerline{\includegraphics[width=0.7\textwidth]{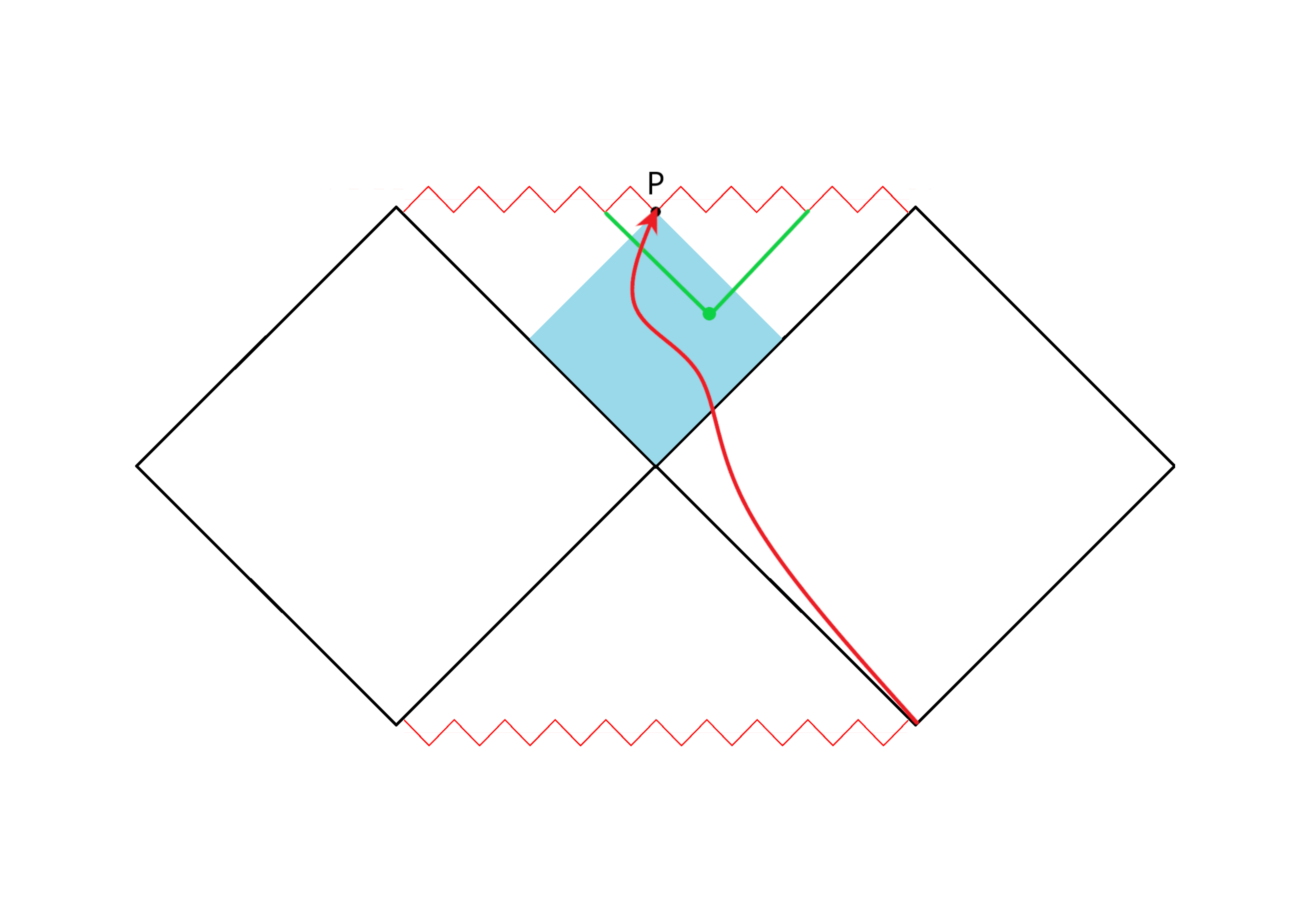}}
\caption{ The blue region is the causal diamond connecting between the horizon bifurcation  and the point where an infalling observer (with a trajectory marked by a red line)  hits the singularity. The infalling observer encounters every IFS produced in the blue region.  The folds of such an IFS are marked by the green lines.    \label{casual diamond}}
\end{figure}

\item {\it  Stringy cosmology} \\

Combining (\ref{jh}) with \cite{Itzhaki:2021scf} we conclude that  in cosmological setups, that involve running dilaton, $\dot\Phi>0$, the effective description of IFSs, at large scales,  is via an  ideal gas with
\be\label{shj}
 P_{IFS}= -e^{-2\Phi} \frac{\dot \Phi^2 }{128 \pi^7 }\tau_{IFS}^2, ~~~~\rho_{IFS}=0,
\ee
where $\tau_{IFS}$ is the life time of an IFS and $P_{IFS}$ and $\rho_{IFS}$ are the pressure and energy density induced by the IFSs.  Since (\ref{shj}) violates the null energy condition in the strongest possible way, it is likely to lead to some fascinating cosmological scenarios. To  explore these scenarios we need to know $\tau_{IFS}$.  There are several interaction  channels that involve IFSs and contribute to $\tau_{IFS}$. For example two IFSs can merge in the way presented in figure  \ref{merge}.

\end{itemize}

The advantage of having a  CFT description of the IFS is that, at least in principle, we  should be able to calculate all the relevant amplitudes and find $\tau_{IFS}$.  Related calculation, in the context of the $c=1$ matrix model, that took advantage of the relevant higher point functions in FZZT-branes  \cite{Ponsot:2001ng,Ponsot:2002ec},   was done in  \cite{Balthazar:2018qdv}.  It would be interesting to analyze various multi-point functions in the presence of IFS to probe more of its dynamical features.
Since the limit in which the regulator is taken to infinity (\ref{bg}) is similar in spirit to the Gross-Mende limit \cite{Gross:1987kza,Gross:1987ar} the calculations are expected to 
simplify in this limit.

\section*{Acknowledgments}

The work of NI is supported in part by a center of excellence supported by the Israel Science
Foundation (grant number 2289/18) and BSF (grant number 2018068).
The work of UP was supported in part by the Alexander Zaks scholarship.
\vspace{10mm}

\begin{figure}
\centerline{\includegraphics[width=0.5\textwidth]{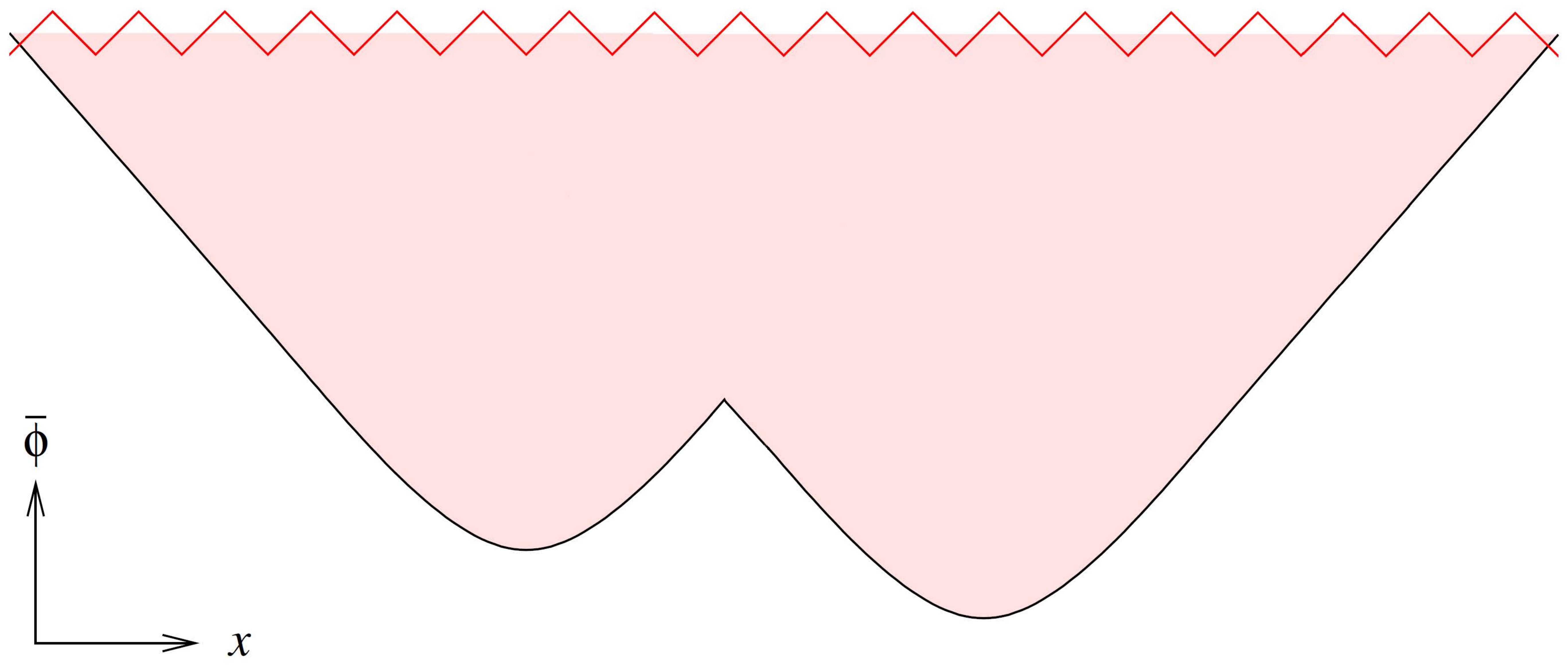}}
\caption{IFS merger is one of the channels that determines $\tau_{IFS}$. \label{merge}}
\end{figure}

\appendix

\section{Detailed computation of FZZT open string two point amplitude \label{appA}}

In this appendix, we will review the technical details for computing the open string two point amplitude $d_\beta(P)$ ordinary Liouville theory and the  continuation to the time-like amplitude $D_{\bar \beta}(\bar P)$. Analysis contrasting the space-like and the time-like setup has been considered previously in \cite{Schomerus:2003vv,Strominger:2003fn,Gutperle:2003xf,Nakayama:2004vk,Zamolodchikov:2005fy,McElgin:2007ak,Harlow:2011ny,Bautista:2021ogd}. The continuation to the time-like setup turns out to be extremely subtle, and the main goal of this appendix is to document these subtleties. These subtleties decouple when we focus on the dynamics of the folded strings. The issue appears to have more to do with the dynamics of open strings on FZZT branes.

\subsection{Action and basic setup}

Let us begin by reviewing the basic construction of FZZT branes in ordinary Liouville theory. The action is
\be S =  \int d^2 \sigma \left({1 \over 4 \pi} (\partial \phi)^2 + {1 \over 4 \pi} Q R \phi +  \mu e^{2b \phi}\right) + \int d \sigma \, \lambda e^{b \phi}, \qquad Q = b + {1 \over b} \ . \label{spaceS} \ee
The boundary vertex operator is
\be B_\beta(x)=  e^{\beta \phi(x)}, \qquad \Delta_\beta = \beta(Q-\beta), \qquad \beta = {Q \over 2} + i P\,  . \label{deltabeta}\ee
A feature which can be seen in the structure of vertex operators is that the dimensions are invariant under exchange $b \leftrightarrow 1/b$. In \cite{Fateev:2000ik,Teschner:2000md}, this property is promoted to a duality principle which will play a crucial role in the computation of the amplitudes as we will review below.

\begin{figure}
\centerline{\includegraphics[width=\hsize]{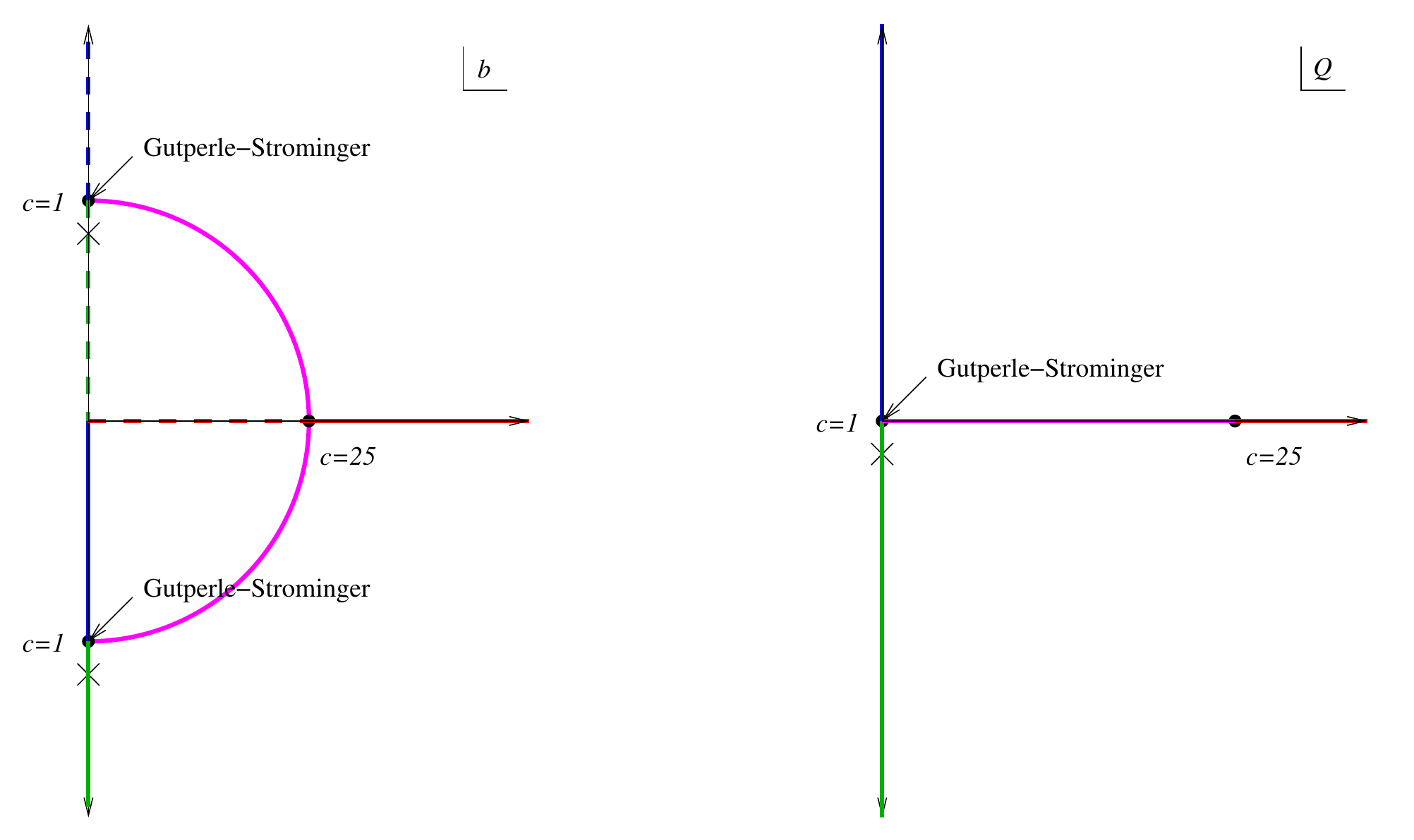}}
\caption{Plane of complex parameters $b$ and $Q$. These two planes are related by two to one map. The two sheets of $Q$ plane embedded in the $b$ plane correspond to the degeneracy of the $b \leftrightarrow 1/b$ exchange. On the $b$ plane, it maps the interior of the pink semi-circle to the exterior.   The $\times$ indicate the point where $\bar b >1$ and its $\bar b \leftrightarrow -1/\bar b$ dual. \label{figa}}
\end{figure}

In order to understand the time-like extension, we will take the continuation (\ref{continue}) giving rise to the action (\ref{TLTaction}).  In order to better visualize this continuation process, it is useful to plot the complex $b$ plane and the complex $Q$ plane parameterizing the space of theories, as is illustrated in figure \ref{figa}. The points along the real $b$ axis corresponds to the ordinary boundary  Liouville setup and the points along the imaginary $b$ axis corresponds to the time-like boundary Liouville setup. The complex $Q$ plane is a double cover of the complex $b$ plane. That there are two values of $b$ corresponding to the same value of $Q$ is a manifestation of the $b \leftrightarrow 1/b$ duality.

\subsection{Computation of two point amplitudes using shift relations}

In this subsection, we will review the computation of the open string two point amplitude
\be d_b(\beta) = \langle B_{\beta}(\sigma) B_{\beta}(\sigma') \rangle (\sigma-\sigma')^{2 \Delta_\beta} \ ,  \ee
using the shift relation following the approach of \cite{Fateev:2000ik,Teschner:2000md}. The purpose of this subsection is to establish notation and to highlight key concepts. We mostly follow the notations and conventions of  \cite{Fateev:2000ik,Teschner:2000md}, and the readers are referred there for a more detailed exposition.

One of the main ingredients in the computation of  \cite{Fateev:2000ik,Teschner:2000md} is the shift relation\footnote{See   \cite{Fateev:2000ik,Teschner:2000md} for the explanation how this shift relation arise.}
\beq \lefteqn{ {d_b(\beta+ b) \over d_b(\beta)}} \cr
&=& - {\pi \over 4 \mu \gamma(1+b^2)}  \Gamma(1-2 b \beta)^{-1} \Gamma (2b \beta - 1)^{-1}
\Gamma(1-b^2 - 2 b \beta)^{-1} \Gamma(2 b \beta - b^2 - 1)^{-1}\cr
&& \times
\prod_{\pm \pm} \sin\left(\pi b {2 \beta \pm i (s_1 \pm s_2)\over 2} \right)^{-1}
 \label{shift}
\eeq
as is explained in  (3.8)-(3.11) of \cite{Fateev:2000ik}. The $s_i$ are defined by
\be \cosh^2 \pi b s_i = {\lambda_i^2 \over \mu} \sin (\pi b^2)  \label{smulambda} \ee
as is given in (2.25) of \cite{Fateev:2000ik} and are to be understood as a parameters which can be used in place of the boundary Liouville coupling $\lambda_i$.

The second important ingredient is the dual shift relation
\beq \lefteqn{{d(\beta + {1 \over b}) \over d(\beta)} }\cr
 &=& - {\pi \over 4 \tilde \mu \gamma(1+{1 \over b^2})} \Gamma\left(1-{2 \beta \over b}\right)^{-1} \Gamma \left({2 \beta \over b} - 1\right)^{-1}
\Gamma \left(1-{1 \over b^2} - {2  \beta \over b}\right)^{-1} \Gamma\left({2  \beta \over b} - {1 \over b^2} - 1\right)^{-1} \cr
&& \times \prod_{\pm \pm} \sin\left({\pi \over b} {2 \beta \pm i (s_1 \pm s_2)\over 2} \right)^{-1}
 \label{dualshift}
\eeq
where
\be \pi \tilde \mu \gamma(1/b^2) = (\pi \mu \gamma(b^2))^{1/b^2} \ee
as is defined in (1.19) of \cite{Fateev:2000ik}. The dual shift relation is a consequence of $b \leftrightarrow 1/b$ duality.

The power of these shift relations is that for strictly real (or strictly imaginary) value of $b$ with irrational $b^2$, the shift relations (\ref{shift}) and (\ref{dualshift}) determine $d(\beta)$ uniquely up to an overall normalization factor, as explained just below (26) of \cite{Teschner:1995yf}. One can then fix $d(\beta)$ uniquely if one further imposes the unitarity relation
\be d(\beta) d(Q-\beta) = 1 \ . \label{unitarity} \ee
The $d(\beta)$ satisfying all these conditions was written in (3.18) of \cite{Fateev:2000ik} as follows
\be d(\beta| s_1,s_2) = (\pi \mu \gamma (b^2) b^{2 - 2 b^2})^{(Q - 2 \beta)/2b} {G (Q-2 \beta) \over G(2 \beta - Q)} \prod_{\pm \pm} S(\beta \pm i (s_1 \pm s_2)/2)^{-1} \ . \label{dbeta} \ee
The functions $G(x)$ and $S(x)$ are as defined in \cite{Fateev:2000ik} will be reviewed more carefully in appendix \ref{app:B}. For now, let us summarize that $d(\beta| s_1,s_2)$ is determined uniquely based on three conditions
\begin{enumerate}
\item shift relation (\ref{shift}),
\item dual shift relation (\ref{dualshift}), and
\item unitarity relation (\ref{unitarity}).
\end{enumerate}

\subsection{Long folded string amplitude in space-like Liouville theory}

In order to keep our analysis from getting needlessly complicated, it is useful to take a limit which simplifies the discussion greatly while maintaining all of the essential physics. This is the limit of vanishing bulk Liouville coupling constant $\mu \rightarrow 0$ while keeping the boundary Liouville constant $\lambda$ finite. We will also assume that there is only one FZZT brane so that
\be s = s_1 = s_2 \ . \ee
Sending $\mu$ to zero then makes $s$ go to infinity. Then, using (\ref{smulambda}), amplitude (\ref{dbeta}) becomes
\beq d_b(\beta) & = & \left({2\pi \lambda b^{(1 -  b^2)} \over \Gamma(1-b^2)}\right)^{-(2\beta-Q)/b}  {G(-2 \beta + Q) \over G(2\beta - Q)} {1 \over S^2(\beta)} \cr
& = &  \left({2\pi \lambda b^{(1 -  b^2)} \over \Gamma(1-b^2)}\right)^{-(2\beta-Q)/b}{\Gamma({2 \beta \over b} - {1 \over b^2}) \Gamma((2 \beta-Q) b)
\over 2 \pi b^{-{2 \beta \over b}+ 2 \beta b + {1 \over b^2} - b^2 - 1}}
  {S(2 \beta) \over S^2(\beta)} \label{dbetaSS} \eeq
where we have used (\ref{Gshift1}) and  (\ref{Gshift2}).
We are interested in setting $\beta = Q/2 + i P$ as we discussed in (\ref{beta}) and explore the large $P$ asymptotic behavior.  To proceed with this analysis, we need to know the asymptotic expansion of $S(\beta)$ for large $P$. This can be inferred from the integral expression (\ref{So1o2}) by isolating the dominant contribution near $t=0$, to write
\be \log S_{\omega_1,\omega_2}(x) \sim \pm \left(-\frac{i \pi  x^2}{2 \omega_1 \omega_2}+\frac{i \pi  x (\omega_1+\omega_2)}{2 \omega_1
   \omega_2}-\frac{i \pi  \left(\omega_1^2+3 \omega_1 \omega_2+\omega_2^2\right)}{12
   \omega_1 \omega_2}+ \ldots \right), \quad \pm (\mbox{Im}\, x) > 0 \ .  \label{SasymW} \ee
When we set $\omega_1 = b$ and $\omega_2 = 1/b$, we recover (A.11) of \cite{Gutperle:2003xf} up to an overall sign.\footnote{There is also a mismatch in the term of order ${\cal O}(x^0)$ which we will not be needing in this analysis. It is likely that this is a typo in (A.11) of \cite{Gutperle:2003xf}. Our (\ref{SasymW}) matches (A.14) of  \cite{Jimbo:1996ss}.} This appears to be traceable to a subtle mismatch in sign conventions outlined in (\ref{FZZJM}).

Combining these ingredients and using Stirling's approximations, we find that
\be
\log d(\beta)= i \pi  P^2+\left(2 i  Q (\log ( 2 bP )-1)-{2 i  \over b}\log \left({2\pi \lambda b^2 \over \Gamma(1-b^2)}\right) \right) P
+{\cal O}  (P^0) \ , \ee
which can be further manipulated to take the form of (\ref{amp}).
The fact that $d(\beta)$ turns out to be a pure phase is a consequence of unitartiy (\ref{unitarity}) when reflecting off a potential wall. The schematic illustration of the scattering open string is illustrated in figure \ref{figc}. In the large $P$ limit, we expect to find that the long folded strings appear as is illustrated there.

\subsection{Analytic continuation to FZZT branes in time-like Liouville theory}

We will now consider the two point amplitude for the open strings on FZZT branes in a time-like Liouville theory. Formally, the action and the vertex operators can be considered as the analytic continuation of their counterparts in the space-like theory.
Specifically, we will take (\ref{continue})
\be b = - i \bar b, \qquad \phi = i \bar \phi \ee
which is the consistent with the standard $i \epsilon$ prescription to relate Minkowski and Euclidean signatures.
The duality relation $b \leftrightarrow 1/b$ and the shift relations (\ref{shift}) and (\ref{dualshift}) also continue smoothly when $b$ is extended to the complex plane.

What is less clear is the status of the solution (\ref{dbeta}) when $b$ is continued to complex values. The uniqueness of solutions which results from the shift and dual shift relations (\ref{shift}) and (\ref{dualshift}) is only guaranteed  either when $b$ is strictly real or when $b$ is strictly imaginary.  It is the case that when $b$ is strictly real (\ref{dbeta}) is well defined and well behaved. One can attempt to approach the amplitude for the time-like theory by either
\begin{enumerate}
\item analytically continuing the solution (\ref{dbeta})  to (\ref{shift}) and  (\ref{dualshift}) to purely imaginary values of $b$ as prescribed in (\ref{continue}) for  a real value of $\bar b$, or
\item find an alternate solution to (\ref{shift}) and (\ref{dualshift}) for the imaginary value of $b$ which is not an analytic continuation of (\ref{dbeta})
\end{enumerate}
The second scenario is possible because uniqueness of (\ref{dbeta}) does not follow when $b$ is not strictly real or strictly imaginary. Infact, for the closed string three point function, this was precisely the situation discussed in  \cite{Zamolodchikov:2005fy} where an alternative solution for the analytically continued $\Upsilon(x)$ function was found. It might turn out to be that it is possible to find an alternate solution via analytic continuation of minimal models along the lines of \cite{Zamolodchikov:2005fy}. We were, however, unable to find an alternate  solution to (\ref{shift}) and (\ref{dualshift}) with sufficient analyticity in terms of $G(x)$ and $S(x)$ to take a large energy and large momentum limit.\footnote{In reference \cite{Bautista:2021ogd}, a scenario similar to 2 was discussed. To the best of our understanding, the amplitude (3.45) of \cite{Bautista:2021ogd} is a solution to the analytically continued (\ref{shift}) but is not a solution to the dual shift relation.}

We will instead pursue the analytic continuation of the the solution (\ref{dbeta}). This approach also turns out to contain various subtleties which we will elaborate further below.

\subsection{Analytic properties of $D_{\bar b}(\bar \beta) = d_{-i \bar b} (i \bar \beta)$}

We will explore the analytic continuation of  $d_b(\beta)$ which we write as
\beq
D_{\bar b}(\bar \beta)&=& d_{-i \bar b} (i \bar \beta)  \label{Dbarbeta}\\
& = &
 (\pi \mu \gamma (-\bar b^2) (-i \bar b)^{2 + 2 \bar b^2})^{-( q - 2  \bar \beta )/2\bar b} {G_{-i \bar b, i/\bar b} (iq-2 i \bar \beta) \over G_{-i \bar b, i/\bar b}(2i \beta - iq)} \prod_{\pm \pm} S_{-i \bar b, i/\bar b}(i (\bar \beta \pm i (\bar s_1 \pm \bar s_2)/2))^{-1}   \ .  \nonumber
\eeq
In order to isolate the contribution of the large instant fold strings, we will be interested in the large $\bar \beta$ asymptotic behavior of this amplitude.  It should be stressed, that (\ref{Dbarbeta}) involves values of $\omega_1$ and $\omega_2$ such that
\be \tau ={\omega_1 \over \omega_2} = b^2 = - \bar b^2 \label{tau}\ee
is real and negative. As was stressed as far back as page 286 of \cite{10.2307/90809} and repeated throughout the literature e.g. \cite{Jimbo:1996ss,Zamolodchikov:2005fy}, the multiple Hurwitz zeta function in terms of which the di-Gamma and di-Sine functions are defined does not converge for this value of $\tau$. We will be taking a closer look at the behavior of these functions as $\tau$ approaches a negative real value.

In order to better understand the behavior of di-Gamma and di-Sine functions under this analytic continuation, we will take a closer look at the integral expressions (\ref{Gint}) and (\ref{So1o2}). The story is somewhat simpler for the case of $S(x)$, so we will discuss that case first, and then we will follow with the discussion of $G(x)$.

Let us parameterize the continuation from positive and real $b^2$ to negative and real $b^2$ by
\be b = e^{-i \theta} \bar b, \qquad \phi = e^{i \theta} \bar \phi \label{continue2} \ee
so that $\theta=0$ corresponds to the space-like case and $\theta=\pi/2$ corresponds to the time-like case  (\ref{continue}). We will take $\bar b$ to be real and positive.  This then corresponds to approaching the imaginary $b$ axis in figure \ref{figa} along a circle centered at the origin in a clockwise direction.

When we substitute
\be \omega_1 = b= e^{-i \theta} \bar b, \qquad \omega_2 = {1 \over b} = {e^{i \theta} \over \bar b}\ee
into (\ref{So1o2}) with the contour of integration fixed, it still converges as long as
\be  |\mbox{Re}(\omega_1 + \omega_2 - 2 x)| < \mbox{Re} (\omega_1 + \omega_2) \ee
or equivalently
\be 0 < \mbox{Re}(x) < \mbox{Re}(\omega_1 + \omega_2) = \bar b \cos\theta + {1 \over \bar b} \cos(\theta) \ . \ee
Since $\bar b>0$, the real part of $\omega_1$ and $\omega_2$ is always positive except for the limiting case of $\theta=\pi/2$. We see that as long as $\theta < \pi/2$, there will be some region of finite size in the complex $x$ plane where the integral (\ref{So1o2}) converges, and as such, we can define it for the entire complex $x$ plane using analytic continuation. The only issue is that the limiting case of interest at $\theta=\pi/2$ is subtle since the domain of convergence in the complex $x$ plane goes to zero in this limit. One can, nonetheless, define $S_{\omega_1,\omega_2}(x)$ for $\theta = \pi/2-\varepsilon$ for the entire $x$ plane by analytic continuation and send $\varepsilon$ to zero at the end.

\begin{figure}
\centerline{\includegraphics[width=3in]{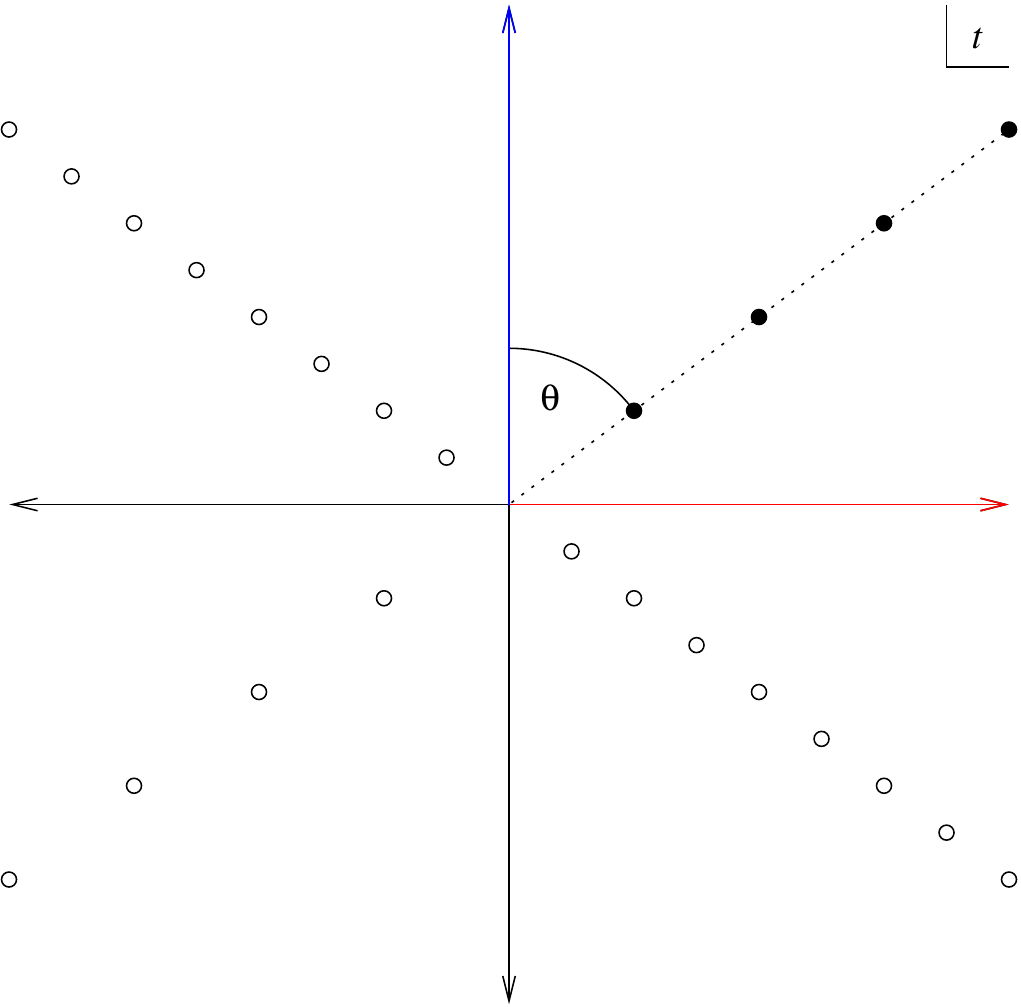}}
\caption{Poles in complex $t$ plane for the integral expression (\ref{So1o2}). Integrating along the positive real axis gives $\log S(x)$. Integrating along the positive imaginary axis gives $\log S^{\bf II}$. The difference between the two is the sum over the residues of poles in the upper right quadrant. \label{figtplane}}
\end{figure}

The integral (\ref{So1o2}) has poles at
\be t = {i \pi m \over \omega_1} = {\pi e^{i (\pi/2+\theta)} \over \bar b} m, \qquad t ={i \pi n \over \omega_2}=  {\pi \bar b e^{i (\pi/2-\theta)}n} \ee
which are illustrated in figure \ref{figtplane}.

In  (\ref{So1o2}), the contour of $t$ integration is taken along the real positive $t$ axis as illustrated in red. We can also evaluate $\log S_{\omega_1,\omega_2}(x)$ as the sum of the integral (\ref{So1o2}) by taking the contour integration to be the positive imaginary $t$ axis illustrated by the blue arrow in figure \ref{figtplane} and the residues of the poles in the $\mbox{Re}\, t > 0$ $\mbox{Im}\, t > 0$ quadrant.  We can then write

\be \log S(x) = \log S^{\bf II}(x) + \Sigma(x) \ee
where $\log S^{\bf II}(x)$ is the quantity obtained by integrating along the positive imaginary $t$ axis, and $\Sigma(x)$ is the sum of residues arising from poles at $t = i \pi n /\omega_2$ with $n \ge 1$.  $S^{\bf II}(x)$ has a finite domain of convergence in the complex $x$ plane in the $\theta \rightarrow \pi/2$ limit.\footnote{What we call $\log(S^{\bf II}(x))$ is essentially what was called $I_b(x)$ in \cite{Gutperle:2003xf}.} As a result, complex features of $S(x)$ at $\theta=\pi/2$ are contained entirely in $\Sigma(x)$.

Some properties of $\Sigma(x)$ can be studied analytically. One can write the sum over residues analytically as follows.
\be \Sigma(x) = \left.  \sum_n {2 \pi i (-1)^n \over   \omega_2 t} {\sinh(\omega_1+\omega_2 - 2x) t \over 2 \sinh( \omega_1 t) } \right|_{t=i \pi n/\omega_2}  \ .  \label{sigmadef} \ee
As was mentioned repeatedly, this quantity was analyzed in \cite{Gutperle:2003xf} and was found to be quite a complicated object. However, it's property under shifts in $\omega_1$ and $\omega_2$ turns out to be much simpler. First let's consider the shift under $\omega_1$. We find
\be \Sigma(x+\omega_1) - \Sigma(x) = -\sum_{n=1}^\infty \frac{2 e^{i \pi  n} \cos \left(\frac{\pi  n (\omega_2-2 x)}{\omega_2}\right)}{n} \ . \ee
This sum can be done as a geometric series by writing
\be \cos x = {e^{i x} + e^{-i x} \over 2} \ee
and the result is
\be \Sigma(x+\omega_1) - \Sigma(x) =  \log \left( 4 \sin^2 \left( {\pi  x \over \omega_2} \right) \right) \ .  \label{sigmashift1}\ee
On the other hand,
\be \Sigma(x+\omega_2) - \Sigma(x) =0  \label{sigmashift2} \ee
is manifest.

This then implies that
\beq S^{\bf II}_{\omega_1 \omega_2}(x+\omega_1) &=& \left(2 \sin\left({\pi x \over \omega_2}\right) \right)^{-1} S^{\bf II}_{\omega_1 \omega_2}(x) \label{SIIshift1} \\
 S^{\bf II}_{\omega_1 \omega_2}(x+\omega_2) &=& 2 \sin \left({\pi x \over \omega_1}\right) S^{\bf II}_{\omega_1 \omega_2}(x) \ .  \label{SIIshift2}\eeq

The function $\exp(\Sigma(x))$ can also be thought of as function defined up to overall multiplicative constant in terms of the shift (\ref{sigmashift1}) and (\ref{sigmashift2}). Some aspects of the structure of this function where studied in  \cite{Gutperle:2003xf}. One feature which is of interest to future applications is the asymptotic behavior for large real values of $x$ (and for (\ref{continue2}) with $\theta = \pi/2$).  For $\mbox{Re}\, x \gg 0$, we find
\beq \Sigma(x) &=& \left.  \sum_{n=1} {2 \pi i (-1)^n \over   \omega_2 t} {\sinh(\omega_1+\omega_2 - 2x) t \over 2 \sinh( \omega_1 t) } \right|_{t=i \pi n/\omega_2} \cr
& = & -\sum_{n=1} {(-1)^n \over n} {e^{(2x - \omega_1 - \omega_2)(\pi n \bar b)} - e^{-(2x - \omega_1 - \omega_2)(\pi n \bar b)}\over e^{\omega_1 \pi n  \bar b} (1 -e^{-{2 \pi \omega_1 n \bar b}}) } \cr
& = &
-\sum_{k=0} \sum_{n=1} {(-1)^n \over n}\left( e^{(2x - \omega_1 - \omega_2)(\pi n \bar b) - \omega_1 \pi n \bar b -{2 \pi \omega_1 kn  \bar b}}-e^{-(2x - \omega_1 - \omega_2)(\pi n \bar b) - \omega_1 \pi n \bar b -{2 \pi \omega_1 kn  \bar b}}\right)
 \cr
& = &  -\sum_{k=0} \log \left(\frac{e^{\pi  \bar b (-2 k \omega_1+\omega_2-2
   x)}+1}{e^{\pi  (-\bar b) (2 (k+1)
   \omega_1+\omega_2-2 x)}+1}\right) \ .  \label{SigmaAsym}
\eeq
The question is whether we can preform the summation over $k$. Actually, the only quantity we really need is
\be F(x) = \Sigma(2 x) - 2 \Sigma(x) \ee
or its derivative since one can reconstruct $F(x)$ from $F'(x)$ up to an additive constant.  It turns out that one can write
\beq  F_K'(x) &=& \cr
&&  \hspace{-.75in} \sum_{k=0}^K   { 2 \pi  \bar b \sin (\pi  \bar b^2 (2 k+1)) \sin (\pi
   (\bar b^2-3 i \bar b x)) \sinh (\pi  \bar b x)\over \sin (\pi
   (\bar b^2 k+i \bar b x)) \sin (\pi  (\bar b^2 k+2 i
   \bar b x)) \sin (\pi  \bar b^2 (k+1)-i \pi  \bar b x)
   \sin (\pi  \bar b^2 (k+1)-2 i \pi  \bar b x)} \cr
   F'(x) & = & \lim_{K \rightarrow \infty} F_K'(x) \ .
\eeq
If $\bar b$ has a small positive imaginary part (corresponding to $b = -i \bar b$ having a small real part), the sum over $k$ truncated at $K$ converges in the large $K$ limit, and in the large $K$ limit, $F'(x)$ appears to grow proportionally to $|x|/\varepsilon$. Clearly, this function does not have a finite $\varepsilon=0$ limit for any values of $\bar b$ in conflict with the claim in \cite{Gutperle:2003xf}. In the strict $\varepsilon=0$ limit, $F'(x)$ appears to be oscillating around a fixed value as a function of $K$. It appears that $F'(x) \approx \mbox{constant}$ is the average large $K$ behavior in the large $x$ limit. As such, it is subleading to the $x^2$ and  $x \log x$, dependent behavior for $x \sim \beta \sim - \bar P + i q/2$ arising from other contributions to (\ref{logD}) and (\ref{logD3}) in the large $\bar P$ limit.

\begin{figure}
\begin{tabular}{ccc}
\includegraphics[width=2in]{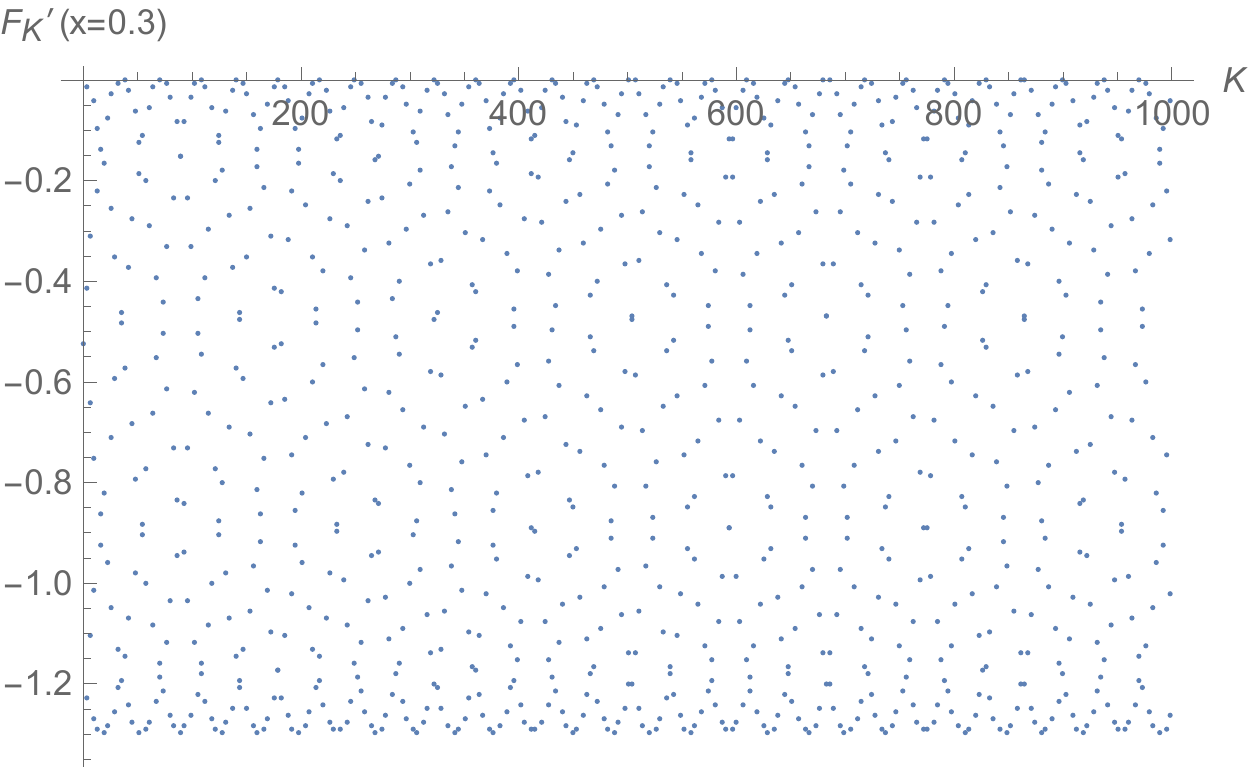} &
\includegraphics[width=2in]{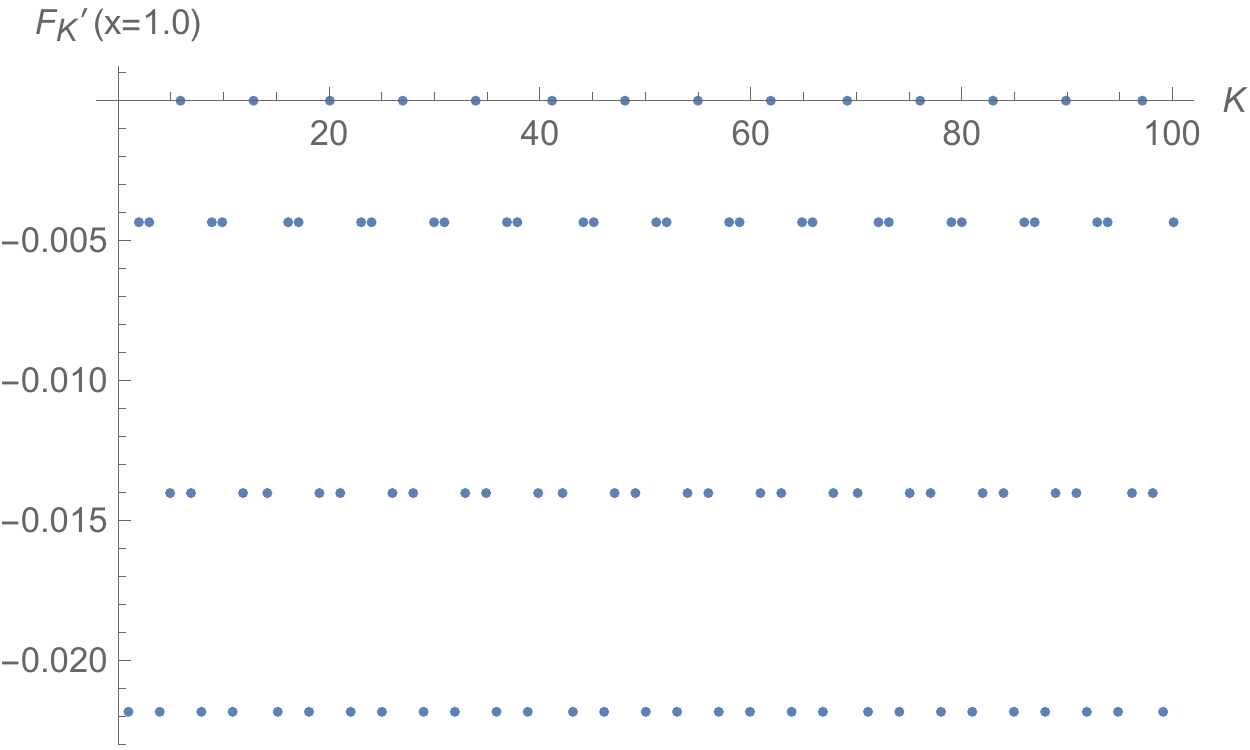} &
\includegraphics[width=2in]{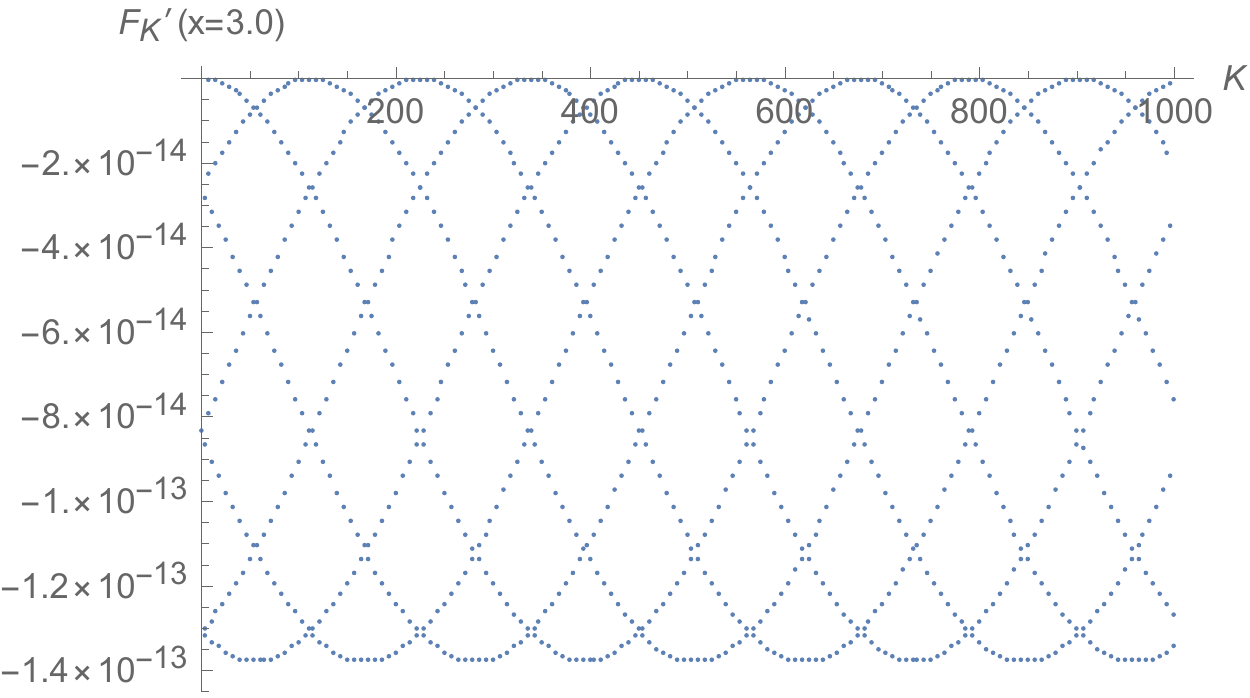} \\
(a) & (b) & (c)
\end{tabular}
\caption{(a) is $\bar b^2 = \sqrt{13}$ and $x = 0.3$. (b) is for $\bar b^2=9/7$ and  $x=1.0$. (c) is for $\bar b^2 = \pi$ and $x=0.3$. \label{fige}}
\end{figure}

To get a sense of the features of $F_K'(x)$, we illustrate the dependence in $K$ for various specific choices of $\bar b$ and $x$ in figure \ref{fige}. These plots are explicitly (semi)periodic, indicating definite structure which we do not fully understand. It appears that for $\bar b^2 = p/q$, $F_K'(x)=0$ for $K = 0\  \mbox{mod}\  q$, and that this structure leads to the semi-periodicity for arbitrary real $\bar b^2$. In addition, for $\bar b = 1$, the sum appears to vanish identically for generic values of $x$. The fact that $F_K'(x)$ is semi-periodic in $K$ implies that it does not converge in the large $K$ limit, but it also implies that it is fluctuating boundedly around some mean value which appears to decrease rapidly as $x$ is increased, as can be seen in the examples illustrated in figure \ref{fige}.

\subsection{Instant folded string amplitude in time-like Liouville theory}

In this section, we will examine the large $\bar P$ limit of the analytically continued amplitude  (\ref{Dbarbeta}). It is actually useful to start with (\ref{dbetaSS}). We will take the continuation
\be \bar \beta = {q \over 2} + i \bar P \ee
so that
\be P = i \bar P, \qquad b = -i \bar b, \qquad Q = i q \label{PbQsubs} \ee
with $\bar P > 0$
which is based on (\ref{continue}).

When $S(x)$ in (\ref{dbetaSS}) is replaced by $e^{\Sigma} S^{{\bf II}}$, the terms linear in $\bar P$ in $\Sigma(2 \beta)-2\Sigma(\beta)$ cancel out, and so $\Sigma(2 \beta)-2\Sigma(\beta) \sim {\cal O}(\bar P^0).$

\begin{figure}[]
\centerline{\includegraphics[]{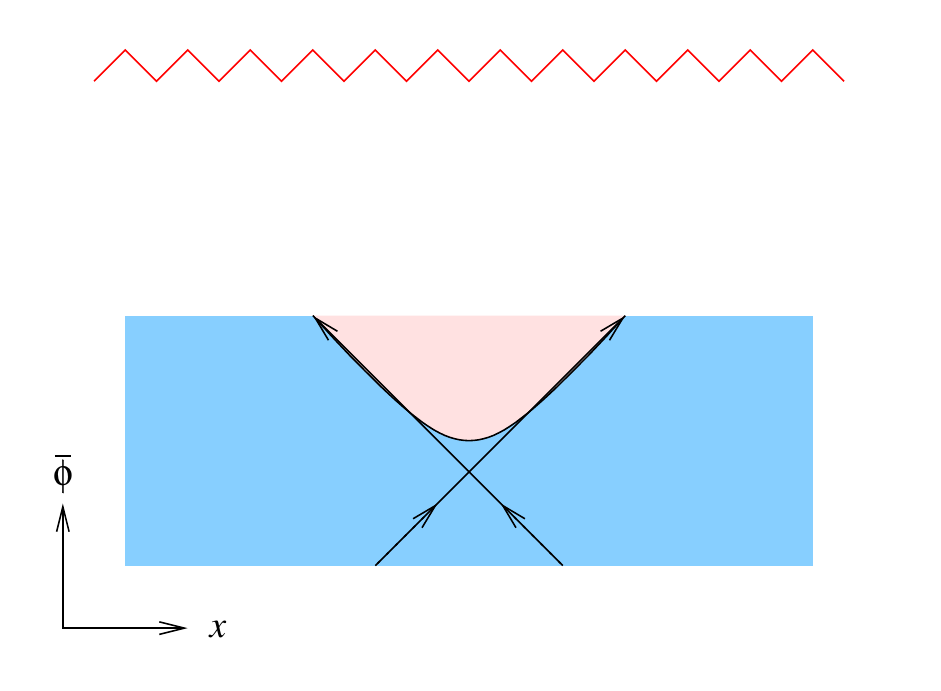}}
\caption{Schematic illustration of timelike FZZT branes and the folded strings with $\bar b > 1$. Since the Liouville field $\bar \phi$ has central charge $c < 1$, additional matter must carry $c > 25$. Here, we are taking $x$ to be a $c=1$ free boson as a part of the $c >25$ matter content, with the remaining $c > 24$ component playing a spectator role. \label{figb}}
\end{figure}

Next, we use the fact that
\beq\lefteqn{\log S^{\bf II}_{-i \bar b,i/\bar b}(i \bar \beta)} \cr
& = & \log S^{\bf II}_{\bar b,-1/\bar b}(-\bar \beta) \cr
&=&   \int_0^\infty {dt \over t} \left[{\sinh(\omega_1+\omega_2 - 2x) t \over 2 \sinh( \omega_1 t) \sinh( \omega_2 t)} - {((\omega_1+\omega_2)/2 - x) \over \omega_1 \omega_2 t} \right]   \label{SIIo1o2} \\
& \sim& \pm \left(-\frac{i \pi  x^2}{2 \omega_1 \omega_2}+\frac{i \pi  x (\omega_1+\omega_2)}{2 \omega_1
   \omega_2}-\frac{i \pi  \left(\omega_1^2+3 \omega_1 \omega_2+\omega_2^2\right)}{12
   \omega_1 \omega_2}+ \ldots \right), \quad \pm (\mbox{Im}\, x) > 0 \nonumber  \eeq
with
\be  x=- \bar \beta \ , \ee
where we have used the scaling relation (This follows from the integral expression (\ref{So1o2}).  See also (B.5) of \cite{Zamolodchikov:2005fy}.)
\be S^{\bf II}_{\omega_1, \omega_2}(x) =S^{\bf II}_{\kappa\omega_1,\kappa \omega_2}(\kappa x) \  . \ee
This leads to the conclusion
\beq \log D_{\bar b}(\bar \beta) & = & \log \left(\langle e^{(-q/2 - i \bar P ) \bar \phi}(x)  e^{(-q/2 - i \bar P) \bar \phi}(y)\rangle (x-y)^{2 \Delta_{\bar \beta}} \right)  \label{logD} \\
& = &
  i \pi  \bar P^2-\left(2 i  q (\log ( 2 \bar b \bar P )-1)-{2 i  \over \bar b}\log \left({-2\pi \lambda \bar b^2 \over \Gamma(1+\bar b^2)}\right) \right) \bar P
+{\cal O}  (\bar P^0) . \nonumber \eeq

The physical interpretation depends on whether we take $0  < \bar b < 1$, or $1 < \bar b$. The case where $1 < \bar b$ is simpler so let us discuss it first. In this case, $q$ is negative, and so the singularity is in the future whereas the FZZT brane is in the past. This leads to the physical configuration illustrated in figure \ref{figb}. The energy is conserved at the edge of FZZT branes.  The case $1< b$ is not compatible with the semi-classical Liouville limit. However, this is separate from the large $\bar P$ Gross-Mende like semi-classical limit. This amplitude is a pure phase if $\lambda$ is taken to be negative. See \cite{Teschner:2003qk} for a related discussion on analytic continuation of $\lambda$.

The same formula, (\ref{logD}), can be applied to the case where $-1 < \bar b < 0$ as is given in (\ref{logD3}) and illustrated in figure \ref{figd}. Note that the range $-1 < \bar b < 0$ is related to the range $1 < \bar b$ by the duality relation $\bar b \leftrightarrow -1/\bar b$.  This is what is reported in (\ref{logD3}).

\subsection{Summary of appendix A \label{appAsummary}}

Since this was a somewhat lengthy appendix, let us summarize the main result here. The open string two point function for FZZT branes in spatial Liouville theory is $d_b(\beta)$ as is shown (\ref{amp}).  The expression (\ref{amp}) was derived as a solution to the shift (\ref{shift}), dual shift (\ref{dualshift}), and unitarity (\ref{unitarity}) relations.

Under analytic continuation (\ref{continue}) and $P \equiv i \bar P$, the action (\ref{TLTaction}) and the shift/unitarity relations (\ref{shift}), (\ref{dualshift}), and (\ref{unitarity}) transform smoothly. At this point, one could either analytically continue the solution and write $D_{\bar b}(\bar \beta) = d_{-i \bar b} (i \bar \beta)$, or attempt to find an alternate solution to  (\ref{shift}), (\ref{dualshift}), and (\ref{unitarity}) after continuation (\ref{continue}) along the lines of
\cite{Zamolodchikov:1995aa}.

We focused our attention on the continuation $D_{\bar b}(\bar \beta) = d_{-i \bar b} (i \bar \beta)$ in the large $\bar P$ expansion. Just like in the spatial case, we found that $\log D_{\bar b}(\bar \beta)$ admits an expansion in large $\bar P$ whose first few leading contributions are of  order $\bar P^2$, $\bar P \log (q \bar P)$, $\bar P^1$, and $\bar P^0$. For $-1 < \bar b < 0$, we obtain the process illustrated in figure \ref{figd} with the amplitude (\ref{logD3}).  We find that the terms of order $\bar P^2$ and $\bar P \log(q \bar P)$ are pure phases. At order $\bar P^1$, the amplitude starts to deviate from being an exactly pure phase, and at order $\bar P^0$, the amplitude is strictly not well defined. This reflects the fact that the di-Gamma and di-Sine functions are not well defined when $\tau$ as  defined in (\ref{tau}) takes on a negative real value.\footnote{The issue is also apparent from the function $\Sigma(x)$ with periodicities  (\ref{sigmashift1}) and (\ref{sigmashift2}) not existing when $\omega_1$ and $\omega_2$ are both real or both imaginary.} We interpret these features at order $\bar P^1$ and $\bar P^0$ to be associated to the pathology of FZZT branes in time-like Liouville theory. It would be interesting to understand this pathology and its possible resolution, which we are leaving for future work. We are interpreting the contributions at order $\bar P^2$ and $\bar P \log (q \bar P)$ as being associated with the creation and propagation of folded strings whose consequences we analyzed in section \ref{sec:rate}.

\section{Aspects of di-Gamma and di-Sine functions \label{app:B}}

In this section, we will review the properties of $G_{\omega_1,\omega_2}(x)$ and $S_{\omega_1,\omega_2}(x)$ for which we are following the notation of \cite{Fateev:2000ik}. It is useful to understand that these functions are closely related to the so called di-Gamma function $\Gamma_{\omega_1,\omega_2}(x)$ and the di-Sine function $S_{\omega_1,\omega_2}(x)$ which where originally defined by Barnes in \cite{10.2307/90809}. A reference frequently cited for technical aspects of the di-Gamma and di-Sine functions is \cite{Jimbo:1996ss}. It is instructive to carefully map out the relation between the conventions of \cite{Fateev:2000ik} and \cite{Jimbo:1996ss} which differ in minor ways which can affect important signs in some of the expressions. Also, in using these functions to study the time-like Liouville theory, we find ourselves pushing various formulas to the limit of their validity.

Let us state for the record that for
\be G^{FZZ}(x) \equiv G_{b,1/b}(x), \qquad S^{FZZ}(x) \equiv S_{b,1/b}(x) \label{FZZJM}\ee
the conventions between  \cite{Fateev:2000ik}, \cite{Jimbo:1996ss}, and  \cite{Bautista:2021ogd} are such that\footnote{The relation between $G^{FZZ}(x)$ of \cite{Fateev:2000ik} and $\Gamma^{JM}_{\omega_1,\omega_2}(x)$ of \cite{Jimbo:1996ss} is also reviewed in Appendix A of \cite{Nakayama:2004vk}.}
\be G_{\omega_1,\omega_2}(x) = {\Gamma^{JM}_{\omega_1,\omega_2}((\omega_1 + \omega_2)/2) \over \Gamma^{JM}_{\omega_1,\omega_2}(x)} = {1 \over \Gamma^{BB}_{\omega_1,\omega_2}(x)} \ee
and\footnote{Because of these subtleties, (A.11) of  \cite{Gutperle:2003xf} is missing an overall minus sign.}
\be S_{\omega_1, \omega_2}(x) =  {1 \over S^{JM}_{\omega_1,\omega_2}(x)}= S^{BB}_{\omega_1,\omega_2}(x) \ .  \ee
One can then write the defining relation for the di-Gamma functions as
\beq G_{\omega_1 \omega_2} (x+\omega_1) &=& {1 \over \sqrt{2 \pi}} \omega_2^{(x/\omega_2 - 1/2)} \Gamma\left({x \over \omega_2}\right) G_{\omega_1 \omega_2}(x) \label{Gshift1}\\
 G_{\omega_1 \omega_2}(x+\omega_2) &=& {1 \over \sqrt{2 \pi}} \omega_1^{(x/\omega_1 - 1/2)} \Gamma\left({x \over \omega_1}\right) G_{\omega_1 \omega_2}(x) \label{Gshift2} \eeq
and
\be S_{\omega_1 \omega_2}(x) = {G_{\omega_1 \omega_2}(\omega_1  + \omega_2 - x) \over G_{\omega_1 \omega_2}(x)} \ee
from which it follows that
\beq S_{\omega_1 \omega_2}(x+\omega_1) &=& 2 \sin\left({\pi x \over \omega_2}\right) S_{\omega_1 \omega_2}(x) \label{Sshift1} \\
 S_{\omega_1 \omega_2}(x+\omega_2) &=& 2 \sin \left({\pi x \over \omega_1}\right) S_{\omega_1 \omega_2}(x) \ .  \label{Sshift2}\eeq
An integral expression can be written for the di-Gamma and the di-Sine functions as follows
\be \log G_{\omega_1,\omega_2}(x) = \int_0^\infty {dt \over t} \left[ {e^{-(\omega_1 + \omega_2) t/2 }- e^{-xt}\over (1 - e^{-\omega_1 t})(1 - e^{- \omega_2 t})} - {((\omega_1 + \omega_2)/2 - x)^2 \over 2} e^{-t} - {(\omega_1 + \omega_2)/2 - x \over t} \right] \label{Gint} \ee
\be \log S_{\omega_1,\omega_2}(x) =  \int_0^\infty {dt \over t} \left[{\sinh(\omega_1+\omega_2 - 2x) t \over 2 \sinh( \omega_1 t) \sinh( \omega_2 t)} - {((\omega_1+\omega_2)/2 - x) \over \omega_1 \omega_2 t} \right]  \ . \label{So1o2}\ee
If we assume that real parts of $\omega_1$ and $\omega_2$ are positive and that this integral converges, one can formally show
(\ref{Gshift1}), (\ref{Gshift2}), (\ref{Sshift1}), and (\ref{Sshift2}) as an identity. Once $G_{\omega_1,\omega_2}(x)$ and $S_{\omega_1,\omega_2}(x)$ are defined for some values of $(\omega_1, \omega_2,x)$, one can consider defining them for all complex values of these variables using analytic continuation.

\bibliography{ifs}\bibliographystyle{utphys}

\end{document}